\documentclass[bibyear,breaklinks=true]{aa}

\usepackage{silence}
\usepackage{graphicx}
\usepackage{lscape}

\usepackage{txfonts}
\usepackage{stackengine}
\usepackage{placeins}

\usepackage{subcaption}
\usepackage{geometry}

\usepackage{float}
\usepackage{tabularx}
\usepackage{longtable}
\usepackage{booktabs}
\usepackage{gensymb}
\usepackage{textgreek}
\usepackage{textcomp}
\usepackage{orcidlink}
\usepackage{booktabs} 
\usepackage{natbib} 

\defcitealias{Kouch2024BlazarNeutrinoAssociation}{K24}
\defcitealias{Plavin2020ApJIntroAGN}{Plavin et al. 2020}
\defcitealias{Hovatta2021A&AIntroAGN}{Hovatta et al. 2021}
\defcitealias{Kouch2025A&AOpticalTracer}{Kouch et al. 2025}

\usepackage{natbib,twoopt}

\bibpunct{(}{)}{;}{a}{}{,}  
\hypersetup{
    colorlinks=true,
    linkcolor=red, 
    citecolor=blue, 
    urlcolor=blue 
}

\begin{document} 

   \title{Probing the statistical correlation of optical tidal disruption events with high-energy neutrinos}

      \author{D.A. Langis \inst{1,2} \orcidlink{0009-0003-9365-9073}
\and
I. Liodakis\inst{1} \orcidlink{0000-0001-9200-4006}
\and
K.I.I. Koljonen\inst{3}\orcidlink{0000-0002-9677-1533}
\and
P.M. Kouch\inst{4,5,6} \orcidlink{0000-0002-9328-2750}
}
\titlerunning{}
\authorrunning{D.A. Langis}
\institute{Institute of Astrophysics, FORTH, N.Plastira 100, Vassilika Vouton, 70013 Heraklion, Greece  \email{dlangis@physics.uoc.gr}
\and 
Department of Physics, University of Crete, 71003 Heraklion, Greece
\and
Institutt for Fysikk, Norwegian University of Science and Technology, Høgskloreringen 5, Trondheim 7491, Norway
\and
Department of Physics and Astronomy, 20014 University of Turku, Finland
\and
Finnish Centre for Astronomy with ESO (FINCA), Quantum, Vesilinnantie 5, 20014 University of Turku, Finland
\and
Aalto University Metsähovi Radio Observatory, Metsähovintie 114, FI-02540 Kylmälä, Finland
}

   \date{Received ; accepted }

  \abstract
   {High-energy (HE) neutrinos have been observed by the IceCube (IC) Neutrino observatory for over a decade. Nevertheless, the astrophysical origin and the responsible mechanisms producing these HE neutrinos are still a mystery, with many astrophysical phenomena as potential emitters. A plethora of previous studies have attempted to study the correlation between HE neutrinos and active galactic nuclei, finding inconclusive results. Tidal disruption events (TDEs) have been proposed as candidate HE neutrino emitters, yet there is only one prior statistical study for the correlation of the two due to the limited number of observed TDEs. For this reason we used TDECat, an optical TDE repository, to investigate the potential association of TDEs with IceCube HE neutrino events. We implemented a spatio-temporal algorithm, where the temporal constraint is based on the transient nature of TDEs. We also simulated two sets of TDEs, correlated differently with neutrinos, to further study their statistical correlation. Despite the individual cases of TDE AT2019dsg and AT2021lo, we find no statistical association between optical TDEs and HE neutrinos. We find jetted TDE Sw J2058+05 to be spatio-temporally associated with a neutrino event. However, a $\gamma$-ray-flaring, flat-spectrum radio quasar is also within the neutrino's sky error region. Although our findings indicate no statistical correlation between optical TDEs and HE neutrinos, this correlation should be further studied in the future. Upcoming surveys such as the Legacy Survey of Space and Time, coupled with next-generation neutrino observatories, such as KM3NET and IceCube-Gen2, will expand both TDE and HE neutrino populations, clarifying their potential correlation.}

   \keywords{
               }

   \maketitle
\nolinenumbers

\section{Introduction}\label{sec:Intro}

High-energy (HE) neutrinos offer a unique glimpse into cosmic accelerators, with observatories such as the IceCube Neutrino observatory \citep{Abbasi2023ApJSIceCat1}, Baikal-GVD \citep{Allakhverdyan2023PhRvDBAIKALGVD}, ANTARES \citep{Fusco2019ICRCANTARES,Albert2024ApJANTARES} and KM3NeT\footnote{\url{https://www.km3net.org/}} \citep{Adri2016JPhGKM3NeT,KM3NeT2025NaturKM3NeT} having detected hundreds in the TeV-PeV range. These HE neutrinos have a chance of being of astrophysical origin, although their sources and the mechanisms responsible for particle acceleration remain a mystery. It has been proposed that accreting supermassive black holes \citep[SMBH; e.g.][]{Mannheim1989A&ABlazarsHadronic,Stecker1991PhRvLAGNneut,Mannheim1992A&AFSRQNEutrinos}, microquasars and/or X-ray binaries \citep[XRBs; e.g.][]{Levinson2001PhRvLXRBneut,Bednarek2005ApJMicroquasars,Christiansen2006PhRvDXRBneut,Koljonen2023MNRASCygX3}, gamma-ray bursts \citep[e.g.][]{Waxman1997PhRvLGRB,Razzaque2003PhRvLGRB, Dermer2003PhRvLGRB,Rudolph2023ApJGRB}, or rapidly spinning neutron stars \citep[e.g.][]{Eichler1978NaturPulsar,Link2005PhRvLPulsars} could produce these HE neutrinos. However, a definitive association has yet to be established, due to the high atmospheric background and the poor spatial resolution of neutrino events. 

Blazars, which are active galaxies whose relativistic jets are directed along our line of sight, are among the most promising candidates for neutrino sources. Due to their jet environment and energetics, blazar jets can create HE neutrinos through both hadronic \citep[e.g.][]{Mannheim1989A&ABlazarsHadronic,Mannheim1992A&AFSRQNEutrinos,Gaisser1995PhRHadrons,Mucke2001APhHadrons} and leptonic processes \citep{Hooper2023PhRvLBlazarsLeptonic}. Some blazars have been spatio-temporally associated with neutrinos, with the most famous being TXS 0506+056 \citep{IceCubeCollaboration2018SciTXS0506+056}. However, past blazar population studies have not convincingly connected blazars with HE neutrinos \citep[e.g.][hereafter \citetalias{Kouch2024BlazarNeutrinoAssociation}]{Plavin2020ApJIntroAGN,Plavin2021ApJIntroAGN,Plavin2024JCAPIntroAGN,Hovatta2021A&AIntroAGN, Kun2022ApJIntroAGN, Suray2024MNRASIntroAGN, Kouch2024BlazarNeutrinoAssociation}. For example, \citet{Plavin2020ApJIntroAGN} used a sample of 3388 jetted active galactic nuclei (AGN) and 56 HE neutrino events (E>200 TeV), and found strong evidence for an association (0.2\% chance probability). In a follow-up study, \citet{Plavin2021ApJIntroAGN} utilized 7 years of IC track data comprising 712,830 detected events, along with 3411 jetted AGN, and found a $3\sigma$ association between the two. Combining their results from both studies, they got a significance of $4.1\sigma$, which was soon challenged by \citet{Zhou2021PhRvDIntroAGN} using the same sample. \citet{Hovatta2021A&AIntroAGN} considered a jetted-AGN dataset observed by the Owens valley radio observatory (OVRO) and Mets{\"a}hori radio observatory with 1795 blazars, which they cross-correlated with a sample of 56 IC neutrino events. Although not all neutrinos were associated with blazars flaring in the radio, they found a $\sim2\sigma$ chance probability when the neutrinos coincided with strong radio flares. Lastly, \citet{Kouch2024BlazarNeutrinoAssociation} investigated the correlation between one of the blazar sub-datasets from \citet{Hovatta2021A&AIntroAGN} and an updated neutrino sample of 283 events, observing a $2.17\sigma$ significance only after enlarging the uncertainty areas by 1$^{\circ}$ in quadrature to account for systematic errors.

Another class of astrophysical phenomena recently proposed as neutrino sources are tidal disruption events \citep[TDEs; e.g.][]{Wang2011PhRvDTDE,Wang2016PhRvDTDE,Stein2021NatAsAT2019dsg, Reusch2022PhRvLAT2019fdr,vanVelzen2024MNRASAT2019aalc,Yuan2024ApJAT2021lwx}. Tidal disruption events occur when a star gets too close to a SMBH and is torn apart by its strong tidal forces. The stellar debris then spirals inwards, forming an accretion disc around the SMBH and emitting a burst of electromagnetic radiation characterized by a fast rise and gradual decay over time \citep{Rees1988NaturTDEs}. The extreme conditions created during such a process (e.g. intense gravitational and magnetic fields, relativistic jets) provide suitable conditions for the acceleration of particles to very high energies. These accelerated particles are theorized to interact with surrounding matter or radiation, producing HE neutrinos \citep[e.g. ][]{Wang2011PhRvDTDE,Lunardini2017PhRvDIntroNeutrinoTDE,Biehl2018NatSRIntroNeutrinoTDE,Hayasaki2019ApJIntroNeutrinoTDE,Murase2020ApJTDE,Wu2024PhRvDIntroNeutrinoTDE}. 

This theorized neutrino production from TDEs took the spotlight when TDE AT2019dsg was found to be spatio-temporally associated with the IceCube neutrino event IC191001A \citep{Stein2021NatAsAT2019dsg}. Subsequently, there have been associations between TDE candidates and neutrino events, namely AT2019aalc - IC191119A \citep{vanVelzen2024MNRASAT2019aalc}, AT2019fdr - IC200530A \citep{Reusch2022PhRvLAT2019fdr} and AT2021lwx - IC220405B \citep[even though it was outside the events reported 90\% error area;][]{Yuan2024ApJAT2021lwx}. Moreover, \citet{Li2024arXivATLAS17jrp} found TDE ATLAS17jrp (AT2017gge) to be spatio-temporally associated with a potential neutrino flare, with a 0.17\% probability of occurring by chance. Furthermore, \citet{Lu2025arXivZTFpop} did a population study on 56 Zwicky Transient Facility (ZTF)-classified TDEs and 143 neutrino events with at least $30\%$ of being astrophysical after 2019, obtaining $\sim1.3\sigma$ significance for the full sample and $\sim2.65\sigma$ for an infrared-selected subsample. 

While the spatio-temporal method has been applied extensively in blazars \citepalias[e.g.][]{Plavin2020ApJIntroAGN,Hovatta2021A&AIntroAGN, Kouch2024BlazarNeutrinoAssociation, Kouch2025A&AOpticalTracer}, this is not the case for TDEs, since there has only been one recent study \citep{Lu2025arXivZTFpop}. Our goal is to test the association between TDEs and neutrinos. For an in-depth review of optical TDEs, readers can refer to \citet{Gezari2021ARA&ATDEs} and \citet{vanVelzen2020SSRvTDEs}.

Our paper is organized as follows: In Sect. \ref{sec:data} we introduce our neutrino and TDE samples , along with their properties. In Sect. \ref{sec:Analysis} we describe our spatio-temporal method as well as the different simulations we conducted to probe its performance for two distinct cases of associated datasets. Finally, in Sect. \ref{sec:results_disc} we discuss our results, followed by the conclusions in Sect. \ref{sec:Conclusions}.

\section{Data}\label{sec:data}

\subsection{TDE sample}\label{ssec:TDEs}

As we mention in Sect. \ref{ssec:neutrinos}, IC neutrino alerts contain detections from $\sim$55056 MJD to $\sim$60232 MJD. We used the optical TDE catalogue (TDECat\footnote{\url{https://github.com/dlangis/TDECat}}) from \citet{Langis2025arXivTDECat} and selected events with durations within the aforementioned time window. Our final sample consists of 107 TDEs (108 flares in total; both flares of AT2020vdq are included). In our sample we do not include events that are not spectroscopically confirmed TDEs \citep[which rules out AT2019fdr, AT2019aalc, and AT2021lwx, that have been suggested to be associated with neutrinos][]{Reusch2022PhRvLAT2019fdr,vanVelzen2024MNRASAT2019aalc,Yuan2024ApJAT2021lwx}.

\begin{figure}[!htbp]
   \resizebox{\hsize}{!}
    {\includegraphics[]{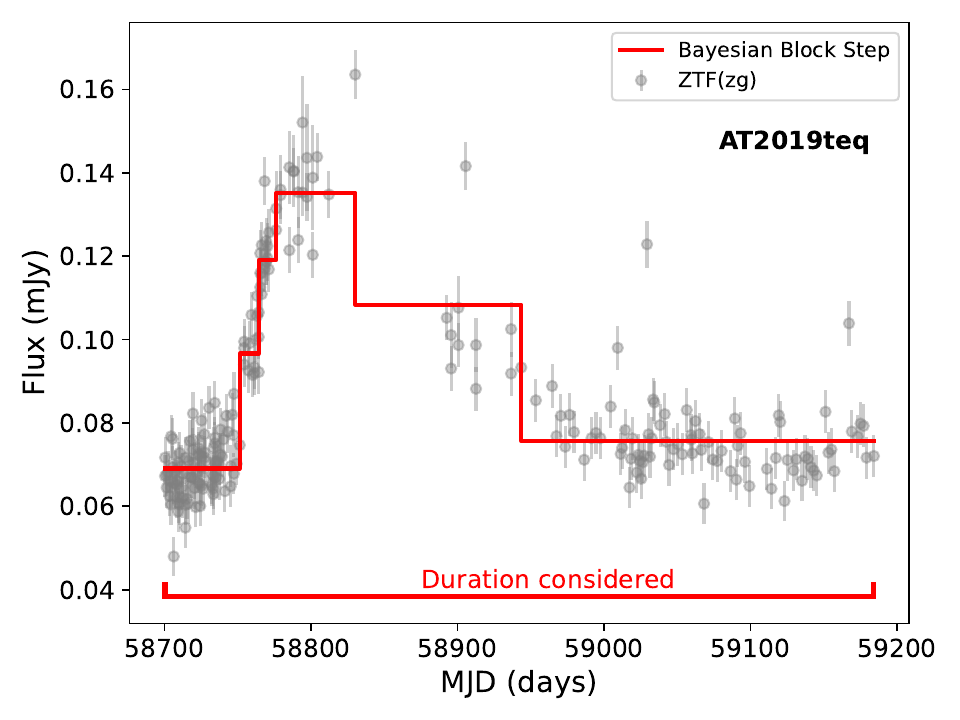}}
    \caption{Example of the Bayesian block algorithm applied to the light curve of TDE AT2019teq. The red bar below the light curve indicates the duration considered in this study.}
    \label{fig:bayesian block}
\end{figure}

To estimate the duration of the events, we utilized the Bayesian block \citep{Scargle1998BayesianBlocks,Scargle2013BayesianBlocks} algorithm from \citet{Langis2025arXivTDECat} (see their Sect. 4.1). We included the uncertainties from the Bayesian blocks in the rise and decay to capture the full duration of the flare, as shown in Fig. \ref{fig:bayesian block}.

However, for eight TDEs (AT2018dyb, AT2018bsi, AT2017eqx, iPTF16axa, iPTF15af, ASASSN-15oi, LSQ12dyw, and PS1-10jh), the flares were not well sampled; hence, the algorithm could not be used. More details about the estimated duration for these sources can be found in Appendix \ref{Appendix:durations}.

\subsection{Neutrino event sample}\label{ssec:neutrinos}

Throughout this study we used the same neutrino sample from \citetalias{Kouch2024BlazarNeutrinoAssociation}, collected using IceCube collaboration neutrino events \citep[IceCat-1;][ except for 8 events since they were induced by cosmic rays]{Abbasi2023ApJSIceCat1}. Additionally, 16 good-quality events were taken from \citet{Abbasi2022ApJK24Sample} which were not included in IceCat-1. The \citetalias{Kouch2024BlazarNeutrinoAssociation} sample includes 283 neutrino events with arrival times between $\sim$55056 MJD to $\sim$59206 MJD (August 2009 to December 2020). Furthermore, we included 73 new alerts added to IceCat-1 as part of a major update\footnote{\url{https://dataverse.harvard.edu/dataset.xhtml?persistentId=doi:10.7910/DVN/SCRUCD}}, which extends the neutrino sample to 356 events and the last arrival time to $\sim60232$ MJD (October 2023). This extended sample is described in more detail and provided as an electronic table in \citet{Kouch2025arXivNeutrinoSample}. It also includes the signalness\footnote{The probability that the neutrino is of astrophysical origin given an astrophysical neutrino spectral index of 2.19$\pm$0.10 and estimated from the expected number of astrophysical neutrinos over the sum of the former and the background at a given declination and energy.} of the events. These, along with the areas of their sky error regions, are used in a neutrino weighting scheme in Sect. \ref{ssec:spatio temporal}.

\section{Analysis}\label{sec:Analysis}

The goal of this work is to investigate whether a potential correlation between optical TDEs and IceCube HE ($\gtrsim100$ TeV) neutrino events exists. We use a similar analysis framework as provided in \citetalias{Kouch2024BlazarNeutrinoAssociation}. This process is described in this section.

\subsection{The spatio-temporal method}\label{ssec:spatio temporal}

Following \citetalias{Kouch2024BlazarNeutrinoAssociation}, we cross-correlated the TDE coordinates with the 90\% likelihood error regions of the neutrinos. We considered a TDE-neutrino event pair to be spatially correlated if the TDE falls within the neutrino error region.

In order to account for the temporal association, we considered a test-statistic (TS) based on the transient nature of TDEs. After estimating the durations of the TDE flares in our sample, we used them for the temporal association. We adopted the activity index (AI) TS from \citetalias{Kouch2024BlazarNeutrinoAssociation}, but modified it to be either 1 or 0, depending on whether the neutrino arrival time falls within the TDE duration or not, respectively. No additional TS was needed, since we studied each individual case after the spatio-temporal associations were found.

After computing the AI TS for each TDE, we summed them globally. This is referred to as the `observed global AI'. Notably, this is equivalent to the total number of spatio-temporally associated TDE-neutrino event pairs. To obtain a null-hypothesis TS value distribution, we chose to randomize the neutrino event RA and the arrival time, while keeping all the other properties the same \citep[e.g.][]{Aartsen2017ApJIceCube}. For each realization (i.e. each randomized neutrino sample), we summed the AI TS globally, obtaining the `randomized global AI'. We repeated this process $10^5$ times, in a Monte Carlo (MC) experiment and retrieved the TS distribution of the randomized global AI values, from which we estimated the chance probability that the TDEs and the IC neutrinos are uncorrelated (i.e. the null hypothesis was rejected). The p-value was found by counting the number of randomized global AI values greater than or equal to the observed global AI value of the dataset (M) over the total number of realizations (N), where \citep{davison1997bootstrap}

\begin{equation}
    p=\frac{M+1}{N+1}
\end{equation}

\begin{figure}[!htbp]
   \resizebox{\hsize}{!}
    {\includegraphics[]{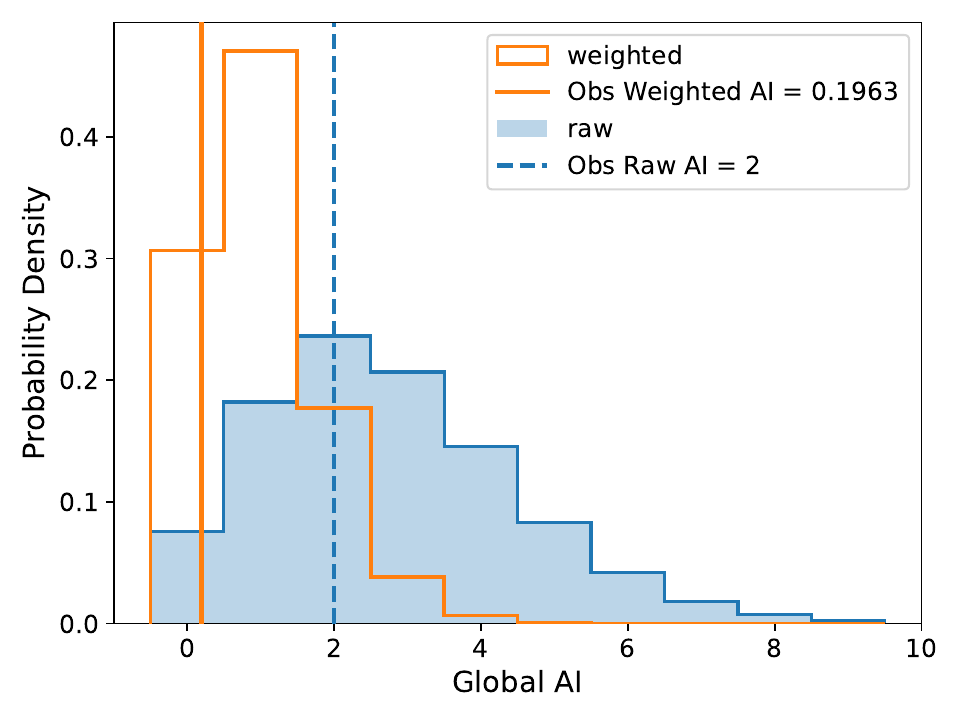}}
    \caption{Raw (sky-blue) and weighted (orange) distributions of the global AI values from the MC experiment, with the respective observed global AI values illustrated with vertical lines.}
    \label{fig:weighted and raw}
\end{figure}

\noindent The distribution of the randomized global AI TS is shown in Fig. \ref{fig:weighted and raw}, where the observed global AI TS is equal to 2 (i.e. only two spatio-temporal associations were retrieved), corresponding to $p_{raw}=0.51$. Hereafter, we refer to this as the 'raw' p-value, since no weights were applied to it. 

We also implemented the same weighting scheme as \citetalias{Kouch2024BlazarNeutrinoAssociation}, utilizing the signalness and the 90\% confidence area as 

\begin{equation}
    W = 
    \begin{cases}
    S, & \text{if } \Omega_{90} \le \widetilde{\Omega}_{90}, \\
    S \cdot \frac{\widetilde{\Omega}_{90}}{\Omega_{90}}, & \text{if } \Omega_{90} > \widetilde{\Omega}_{90}.
    \end{cases}
\end{equation}

\noindent where \(\Omega_{90} = \frac{\pi}{4} \bigl(\alpha^+ \cdot \delta^+ + \alpha^- \cdot \delta^+ + \alpha^- \cdot \delta^- + \alpha^+ \cdot \delta^-\bigr)\) is the 90\% confidence area in the sky ($\alpha$ and $\beta$ correspond to RA and DEC respectively), $S$ is the signalness and $\widetilde{\Omega}_{90}$ (6.73 deg$^2$ for our neutrino event sample) is the global median of the area of the $90\%$ error region for the whole HE neutrino sample. Using these weights, poorly reconstructed events could be included without significantly affecting the overall statistics (for more information see Sect. 3.2 of \citetalias{Kouch2024BlazarNeutrinoAssociation}). The upper panel of Fig. 4 in \citetalias{Kouch2024BlazarNeutrinoAssociation} demonstrates a heat map of how the weights (likewise corresponding to the weighted AI values for each neutrino event) change with $\Omega_{90}$ and $S$.

We repeated the same process used to obtain the raw p-value, but now with the weights. The observed global weighted AI TS is 0.1963, while the distribution of the randomized, weighted global AI TS is shown in Fig. \ref{fig:weighted and raw} (orange), yielding $p_{weighted}=0.85$. We note that these p-values are pre-trial. 

We were alerted about the IceCat1 update after performing the analysis; therefore, we accounted for two trials in both the raw and weighted p-values. We performed a similar analysis to that of \citetalias{Kouch2024BlazarNeutrinoAssociation}, who applied trial corrections to account for the look-elsewhere effect, using the harmonic mean p-value from \citet{Wilson2019PNAStrials}:

\begin{align}
    &\notag\mathring{p}_{raw} = \frac{\sum^L_{i=1}w_i}{\sum^L_{i=1}w_i/p_{i,raw}}=0.584,\\ &\mathring{p}_{weighted} = \frac{\sum^L_{i=1}w_i}{\sum^L_{i=1}w_i/p_{i,weighted}}=0.712
\end{align}

\noindent where L is the number of p-values for each case (two for each in our case), $p_i$ corresponds to each individual p-value, and $w_i$ is the weight for each p-value (sum of the weights should be equal to 1). Since the setup was the same, the weights were the same, $w_i=1/L=0.5$. To test the null hypothesis at a predefined significance level (using a false-positive rate of $\alpha$) using $\mathring{p}$, the threshold $\alpha$ was adjusted to account for the total number of hypotheses tested (L=2 in our case). We used a threshold corresponding to $2\sigma$, which yields a false-positive rate of $a_{2\sigma}=0.0455$. Additionally, we used Eq. 4 from \citet{Wilson2019PNAStrials} to calculate the adjusted $\alpha_{2\sigma|L=2}\approx 0.03899$, which we then compared with the raw and weighted cases - i.e. $\mathring{p}_{raw}\approx0.584>0.03899\approx\alpha_{2\sigma|L=2}$ and $\mathring{p}_{weighted}\approx0.712>0.03899\approx\alpha_{2\sigma|L=2}$ respectively. Since both $\mathring{p}_{raw}$ and $\mathring{p}_{weighted}$ do not satisfy the $2\sigma$ threshold, we conclude that there is no statistical correlation between our optical TDE and neutrino samples.

\subsection{Simulations}\label{ssec:simulations}

We performed simulations for two cases of generated mock datasets: 1) completely correlated neutrino-TDE data and 2) completely random neutrino-TDE data. For the former, we generated TDE data that are both spatially and temporally coincident with the neutrino sample, whereas for the latter, the TDE data were randomly generated. In all simulations we considered the observed neutrino data, since IC sensitivity depends on the declination angle \citep{Abbasi2023ApJDecDep}, meaning that randomly generating them would introduce inconsistencies. Additionally, we performed a series of simulations to attain the number of associations required for statistical association.

\subsubsection{Correlated data}\label{sssec:correlated}

We generated correlated data that simulate our real data. We set 108 TDE flares, consistent with our real data, ranging from 55056 to 60232 MJD. First, we generated the TDE flare ranges by drawing common logarithm TDE decay times from a normal distribution of $\mu = 2.39$ and $\sigma= 0.29$. In \citet{Langis2025arXivTDECat} we found that the variability timescales of TDEs are consistent with log-normal distributions, where the aforementioned $\mu$ and $\sigma$ describe the decay duration of the confirmed TDE sample. In addition, we used the orthogonal distance regression (ODR) fit results from \citet{Langis2025arXivTDECat} for $\log_{10}(t_{rise})$ versus $\log_{10}(t_{decay})$, where $\log_{10}(t_{rise}) = \alpha \log_{10}(t_{decay})+\beta$, $\alpha = 0.915 \pm 0.075$ and $\beta = -0.31 \pm 0.17$, to retrieve the corresponding rise times. We then added the rise and decay times to get the duration of the TDE flare. We thus computed these rather than just drawing them from the duration log-normal distribution for reasons given below. The TDE coordinates were generated as described below.

For the neutrino events, we first determined those from the observed sample that are astrophysical using the signalness $S$ of all 356 neutrinos. For each neutrino's signalness, we drew a random number $x$ from a uniform distribution between $[0,\ 1)$. If $S>x$, we then considered that neutrino to be astrophysical. Next, we randomly selected 108 of them and assigned each a different TDE, that is both spatially and temporally coincident (i.e. the TDE coordinates are inside the error region of the neutrino and the neutrino arrival time falls within the flare duration ). The other 248 neutrinos were assigned random arrival times and coordinates. This resulted in a sample where 108 TDE flares are spatio-temporally associated with (at least) 108 of the 356 neutrinos.

When we applied both the raw and weighted spatio-temporal tests to these 108 associated TDE-neutrino pairs, we retrieved a raw global AI TS equal to 108 (or 109/110 or higher for additional random associations). As expected, these tests resulted in $p=10^{-5}$ (the lowest possible p-value given our MC count of $10^5$), indicating maximal correlation between the simulated TDEs and neutrinos. This process was repeated for 1000 different correlated datasets to obtain a p-value distribution. As expected, for all the 1000 different correlated datasets, the p-value is the same.

\begin{figure}[!htbp]
   \resizebox{\hsize}{!}
    {\includegraphics[]{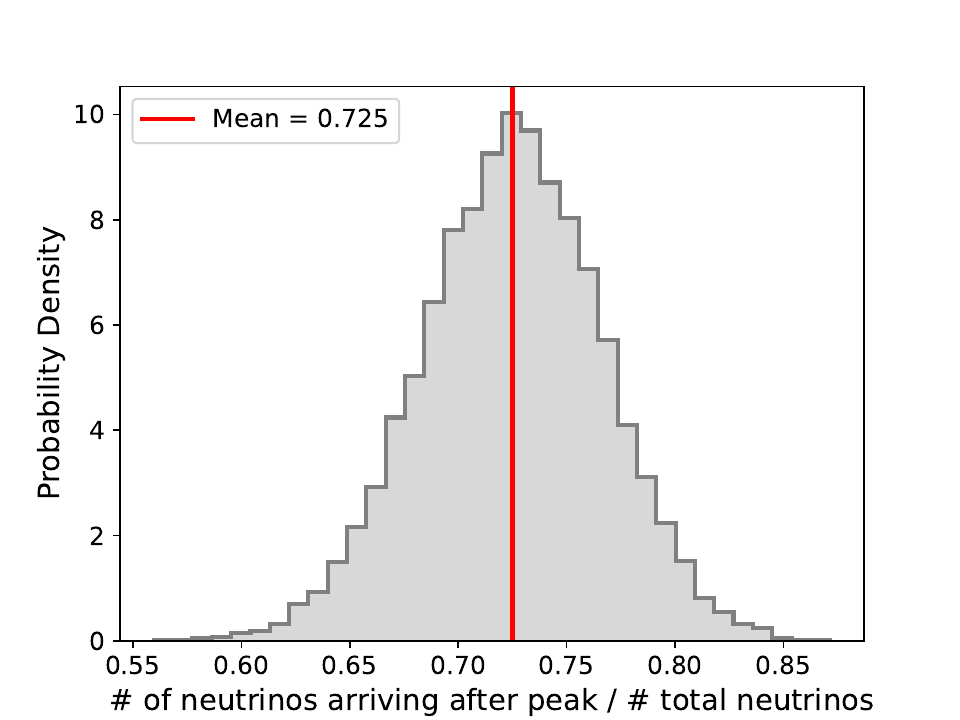}}
    \caption{Distribution from the simulations of the number of neutrinos arriving after TDE peaks over the total number of neutrinos (see Sect. \ref{sssec:correlated}).}
    \label{fig:neutrino arrival}
\end{figure}

However, for these 1000 different datasets, we also examined whether the neutrino arrived before or after the peak by chance, by counting the number of neutrinos arriving during the decay phase for each correlated dataset. Figure \ref{fig:neutrino arrival} shows the distribution of neutrinos arriving during the TDE flare decay time relative to the total number. The mean of the distribution is $0.72\pm0.04$ (where the error is the standard deviation), indicating that due to the asymmetry between the rise and decay times, 72\% of the neutrinos in our simulations arrive after the peak.

\begin{figure*}[!htbp]
   \resizebox{\hsize}{!}
    {\includegraphics[]{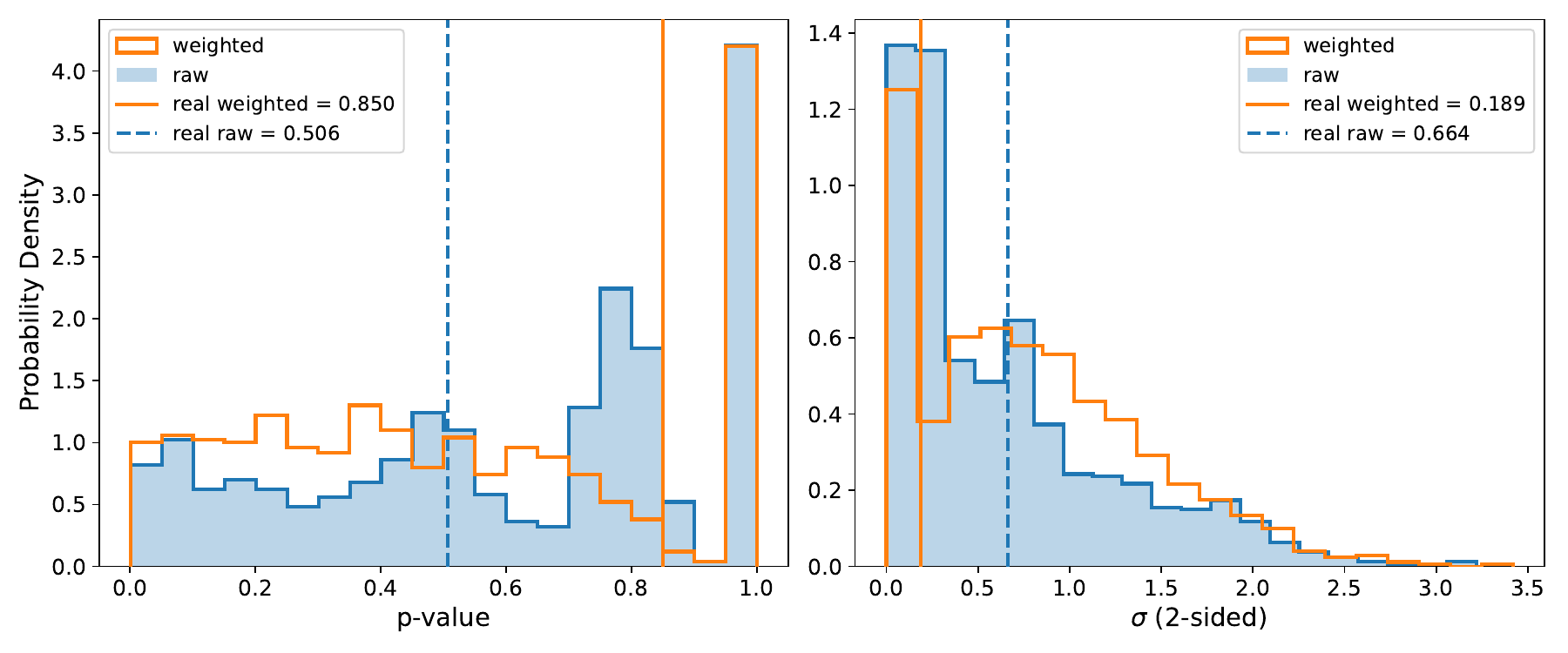}}
    \caption{Raw (sky-blue) and weighted (orange) p-value and sigma-level distributions from the random data simulations of 1000 generated datasets (see Sect. \ref{sssec:random}). The vertical lines indicate the real values raw (blue) and weighted (orange) values from the observed data.}
    \label{fig:sim pval sigma}
\end{figure*}

\subsubsection{Random data}\label{sssec:random}

We followed the same process as Sect. \ref{sssec:correlated} to generate our artificial data, but with TDE RA and DEC drawn randomly from uniform distributions. We then applied the spatio-temporal method to retrieve the randomized global AI distributions and p-values for both the raw and weighted cases. We repeated it for 1000 different random datasets to obtain the p-value distributions. Both the raw and weighted distributions for the p-values and the Gaussian sigma\footnote{$\sigma = \sqrt{2}erf(p)$} are shown in Fig. \ref{fig:sim pval sigma}. We note that the raw distribution (sky-blue) is bimodal due to low-number statistics, given the zero-association threshold in some of the realizations.

\subsubsection{Required associations for signal}\label{sssec:signal}

We used the same process as described in Sect. \ref{sssec:correlated}, but now for each 1000 datasets generated, we decrease the number of correlated TDEs by one (we start with at least 108 correlated TDEs-neutrinos and end with completely random data, i.e. Fig. \ref{fig:sim pval sigma}). In the end, we have 109 different p-value distributions (both weighted and raw), with different simulated association numbers. We note that the total number of associations is larger or equal to the simulated association number, since there are a few that occur due to random chance in each dataset.

\begin{figure}[ht]
   \resizebox{\hsize}{!}
    {\includegraphics[]{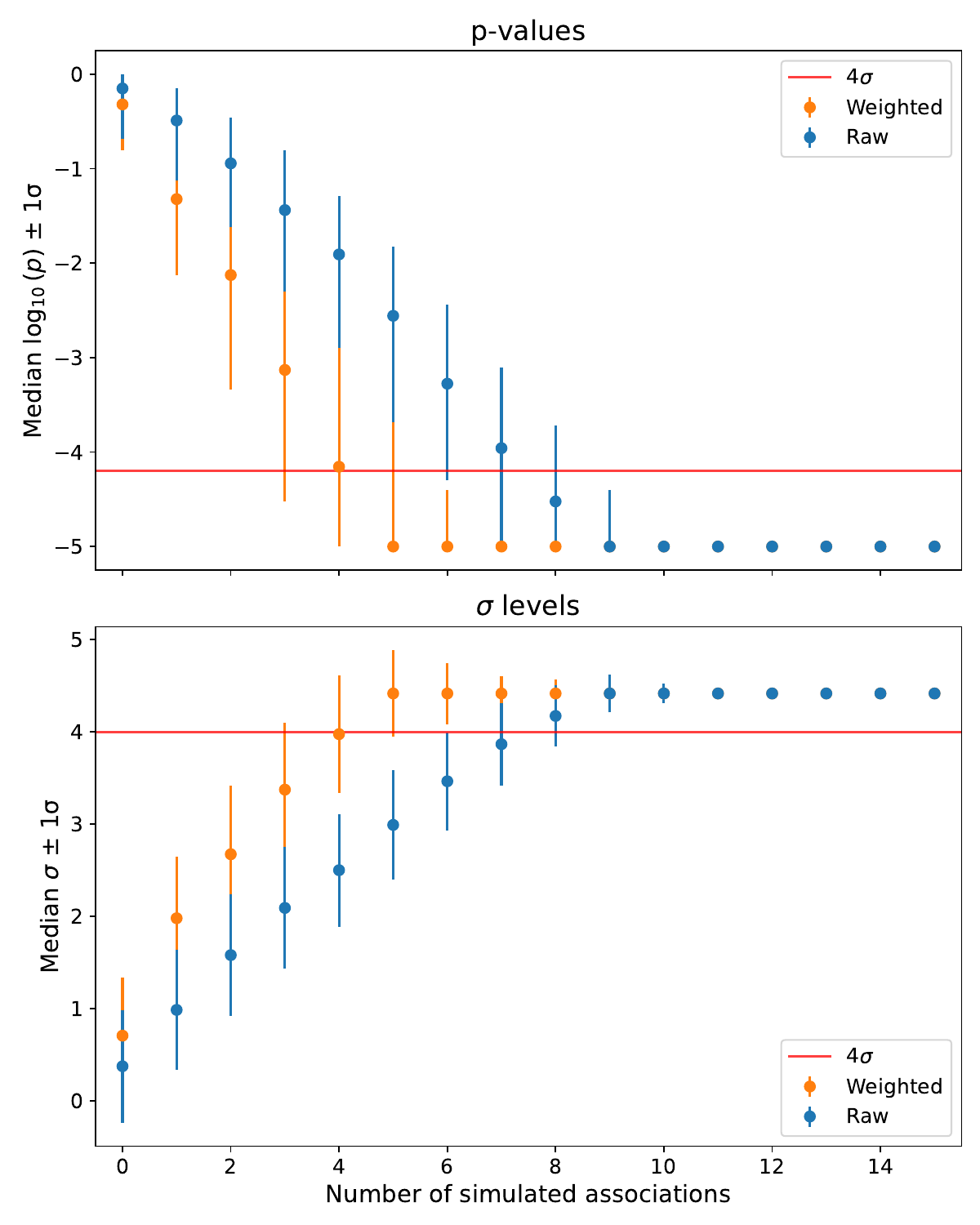}}
    \caption{Median p-values and corresponding sigma levels for the weighted (orange) and raw (sky-blue) cases versus the number of simulated associations. The horizontal red lines indicate the p-value equivalent of $4\sigma$ (upper panel) and the $4\sigma$ threshold (lower panel).}
    \label{fig:signal}
\end{figure}

The upper panel of Fig. \ref{fig:signal} shows the median p-values as a function of the simulated association number for both the raw (sky-blue) and weighted (orange) cases. For the positive error we consider the distance of the 84$^{th}$ quantile from the median, while for the negative error the distance from the median to the 16$^{th}$ quantile. In the bottom panel, the corresponding median $\sigma$ levels are again shown as a function of the simulated association number.

We observe that in order to get a 4$\sigma$ detection (i.e. probability is more than 1$\sigma$ of the 4$\sigma$ threshold), we need 6 simulated associations (i.e. at least 6 TDEs need to be associated with astrophysical neutrinos in our sample). In the raw case, we need at least 9 simulated associations.

\section{Results and discussion}\label{sec:results_disc}

While a statistical association between neutrinos and our optical TDE sample is not significant, it is still interesting to examine individual sources that could potentially be neutrino emitters. Through our cross-correlation we found that only TDE AT2019dsg and TDE AT2021lo are spatio-temporally associated with neutrino events, IC191001A and IC220304A. The former has been reported in the literature \citep{Stein2021NatAsAT2019dsg}, where the neutrino arrives after the optical peak and is roughly coincident with the infrared peak, as seen in the upper panel of Fig. \ref{fig:blazar}. For AT2021lo, we find that the neutrino arrives after the optical peak as well; however, it displays no infrared variability and has no X-ray or ultraviolet publicly available data. Since the mechanisms that could produce neutrinos in the context of TDEs are still unclear and since it is more likely for the neutrino to arrive after the optical peak by chance (Sect. \ref{sssec:correlated}), we cannot confirm if it is purely by chance, even though there have been proposed mechanisms that would account for the delay with respect to the optical peak \citep[e.g.][]{Winter2021NatAsdelayed}.  

To get a simple estimate for the fraction of the TDE diffuse neutrino flux, $f_{TDE}$, we utilized an analysis similar to that used by \citet[][see their Sect. 2]{Bartos2021ApJPieChart}. We obtained 

\begin{equation}\label{eq:fraction} 
    f_{TDE,>2018}\ (f_{TDE,<2018}) =\frac{N_{det}}{N_{neutr}Sf_{C,>2018 \ or\ <2018}}
\end{equation}

\noindent where $N_{det}=2$ is the number of detections, $N_{neutr}=356$ is the total number of neutrino events in our sample, $S=0.42$ is the median signalness of our sample and $f_C$ is a completeness factor for our TDE sample. The TDEs in our sample exhibit optical flares that start between 2008 and 2023. The Zwicky Transient Facility \citep[ZTF;][]{Graham2019ZTFScObj, Bellm2019ZTFSys, Masci2019ZTFDataPr} started observing TDEs from 2018, with its extreme field of view and high cadence. For this reason, the completeness factor of our TDE sample was split into two different time periods, one before 2018 and one after. For the former, we adopted a conservative $f_{C,<2018}=0.1$ to avoid overestimating $f_{TDE}$, and for the latter we considered $f_{C,>2018}=0.5$ \citep[same as][]{Bartos2021ApJPieChart}. Hence, from Eq. \ref{eq:fraction} we obtain $f_{TDE,>2018}=2.7\%$ and $f_{TDE,<2018}=13.4\%$. Even considering $N_{det,\ raw}=9$ or $N_{det,\ weighted}=6$ for the TDEs observed after 2018, which are the required associations in order to have a clean $4\sigma$ statistical association of TDEs with neutrinos for the raw and weighted schemes respectively (see Sect. \ref{sssec:signal}), we retrieve $f_{TDE,\ raw}\approx 0.12=12\%$ and $f_{TDE,\ weighted}\approx 0.08=8\%$. Despite their limited robustness, these results are consistent with \citet{Stein2019ICRCrefereepaper}, who found that non-jetted TDEs are responsible for less than 26\% of the astrophysical neutrino flux. We note that the completeness parameter is highly uncertain; hence, these estimates should be treated with caution. A detailed calculation would need to account for different TDE population models; such a calculation lies beyond the scope of this work.

\subsection{TDE AT2019dsg/IC191001A}\label{ssec:AT2019dsg/IC191001A}

The neutrino event IC191001A has a signalness of 0.59 (more likely astrophysical) and has $\Omega_{90}=24.29$ deg$^2$.  
Naively, HE neutrinos are expected to be accompanied by $\gamma$-ray emission, since their hadronic production channel ($p\gamma$ and/or $pp$) also generates neutral pions that decay into $\gamma$ rays. For this reason, we repeated the spatial association part of the algorithm for the sources in the Large Area Telescope (LAT) 14-year Source Catalog\footnote{\url{https://fermi.gsfc.nasa.gov/ssc/data/access/lat/14yr_catalog/}} \citep[4FGL:][]{Ballet20234FGL,Abdollahi2022LAT} and neutrino event IC191001A. We found that three 4FGL sources are spatially coincident with IC191001A, namely 4FGL J2052.7+1218, 4FGL J2117.0+1344 and 4FGL J2115.2+1218. The former two are millisecond pulsars (MSPs) and the latter is a blazar candidate. 

Pulsars have been linked with HE neutrino emission since the late 1970s \citep[e.g.][]{Eichler1978NaturPulsar,Bednarek1997PhRvLCrabPulsar} and have continued to be studied \citep[e.g.][]{Link2005PhRvLPulsars,Link2006MNRASPulsars,Bhadra2009MNRASPulsars, Dey2021BrJPhMSP}. In particular, \citet{Dey2021BrJPhMSP} studied newborn MSPs as possible candidates for HE neutrino sources, where they focused on a purely leptonic scenario. They studied studied the Langmuir-Landau-Centrifugal-Drive (LLCD) mechanism for accelerating electrons leading to PeV neutrinos and $\gamma$ rays. Moreover, \citet{Dey2021BrJPhMSP} find that the observed $\gamma$ rays would be at TeV due to stronger absorption in their production region and the long distance travelled to reach our telescopes. The 4FGL catalogue spans 50 MeV to 1 TeV but is optimized for the GeV range; hence, we do not observe TeV $\gamma$ rays from the two aforementioned MSP sources, 4FGL J2052.7+1218 and 4FGL J2117.0+1344.

\begin{figure}[!htbp]
   \resizebox{\hsize}{!}
    {\includegraphics[]{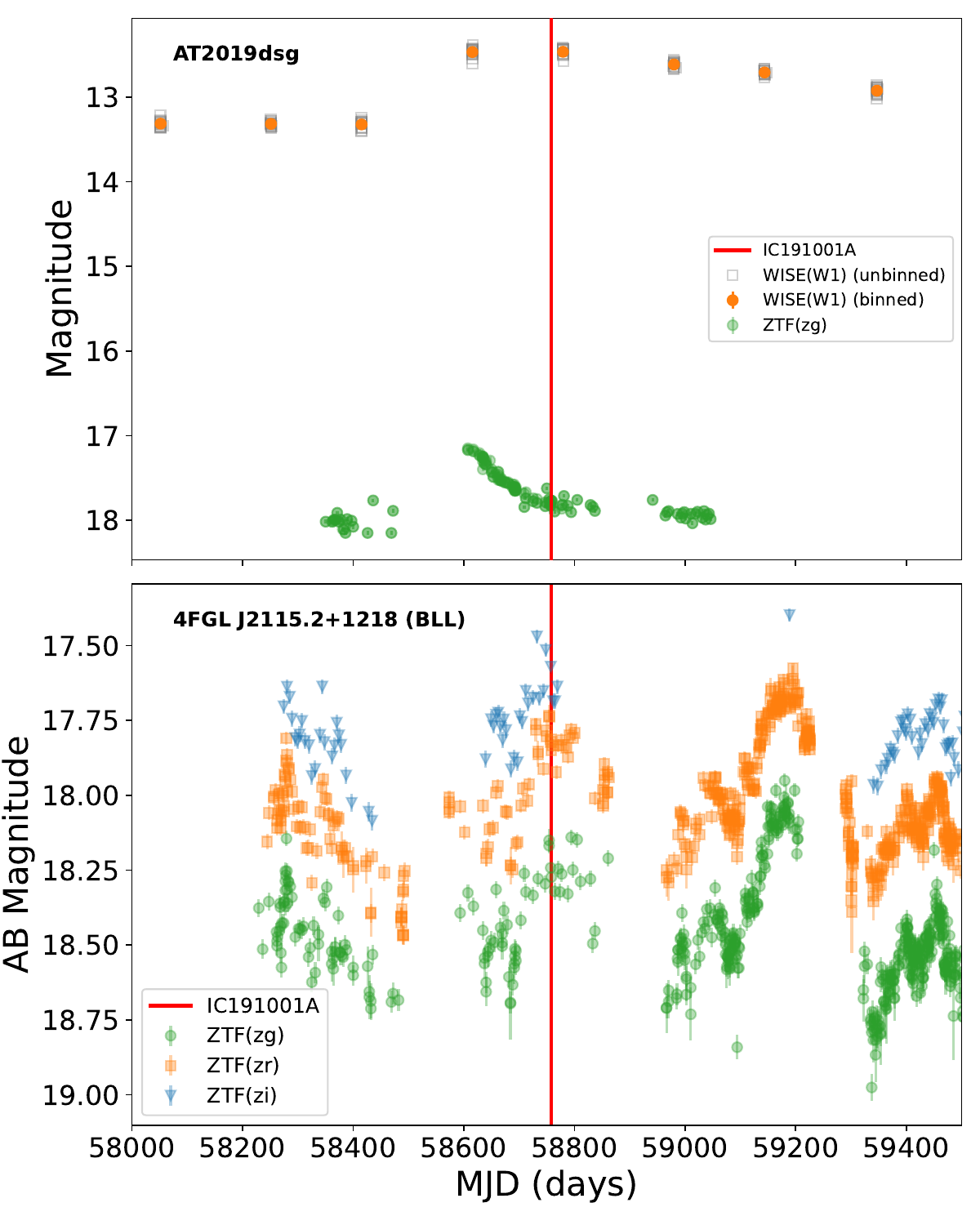}}
    \caption{Light curves for TDE AT2019dsg (optical and infrared; upper panel) and 4FGL J2115.2+1218 (optical; bottom panel). The vertical red line indicates the neutrino arrival time of IC19001A.}
    \label{fig:blazar}
\end{figure}

The blazar candidate spatially coincident with IC191001A, 4FGL J2115.2+1218, was in a flaring state in the optical at the time of the neutrino arrival, as seen in the bottom panel of Fig. \ref{fig:blazar}. In contrast, the yearly binned 4FGL light curve shows only 2$\sigma$ upper limits during that period. Although $\gamma$-ray and optical emission in blazars are typically correlated \citep[e.g.][]{Liodakis2019ApJopticalgammablazar}, models exist that can explain their absence in the context of neutrino production. \citet{Murase2016PhRvLHiddenCR} proposed a photohadronic model for the production of the IC diffuse neutrino flux, where neutrino production is observed without the source being detected in the $\gamma$ rays due to it being opaque to the $\pi^0$-decay $\gamma$ rays via $\gamma\gamma\rightarrow e^+e^-$ cascades. There has also been observational evidence for the potential association of $\gamma$-ray-suppressed blazars with neutrinos \citep{Kun2021ApJDarkBlazars, Kun2023A&ADarkBlazars}. Although it remains uncertain whether optical activity is a good tracer of neutrino activity (\citetalias{Kouch2024BlazarNeutrinoAssociation}; \citealt{Kouch2025A&AOpticalTracer}), this optical-flaring, $\gamma$-ray dark blazar candidate could be responsible for the production of neutrino event IC191001A, instead of TDE AT2019dsg.

\subsection{TDE AT2021lo/IC220304A}

\begin{figure}[!htbp]
   \resizebox{\hsize}{!}
    {\includegraphics[]{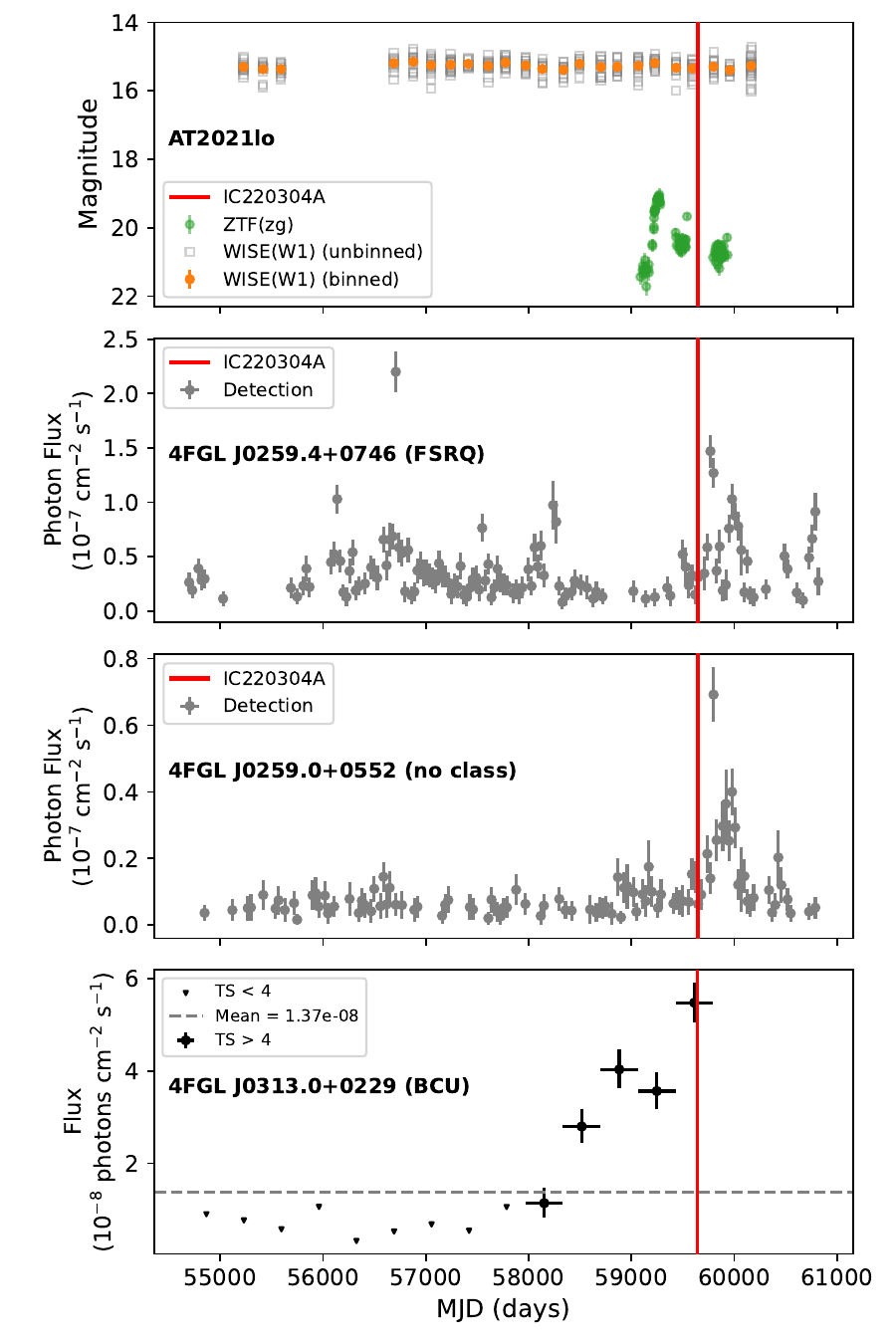}}
    \caption{Light curves for TDE AT2021lo (optical and infrared; first panel), 4FGL J0259.4+0746 ($\gamma$-ray observations in monthly bins; second panel), 4FGL J0259.0+0552 ($\gamma$-ray observations in monthly bins; third panel) and 4FGL J0313.0+0229 ($\gamma$-ray observations in yearly bins; final panel). The red vertical line indicates the neutrino arrival time of IC220304A.}
    \label{fig:AT2021lo_4FGL}
\end{figure}

The neutrino event IC220304A has a signalness of 0.631 (more likely astrophysical) and exhibits $\Omega_{90}=128.91$ deg$^2$. We repeated the spatial association with the 4FGL sources and the error region of neutrino event IC220304A (like in Sect. \ref{ssec:AT2019dsg/IC191001A}). We find that 20 4FGL sources are coincident with IC220304A. These include eight BL Lacs (BLL), five blazars of uncertain type (BCU), two flat-spectrum radio quasars, one radio galaxy (RDG), one MSP and three unclassified sources. Out of these sources, three seem to flare in the $\gamma$ rays around the neutrino arrival time (59642.74 MJD), namely 4FGL J0259.4+0746 (FSRQ), 4FGL J0259.0+0552 (not classified), and 4FGL J0313.0+0229 (BCU). The $\gamma$-ray light curves for the aforementioned sources, along with the optical and IR light curve of AT2021lo are shown in Fig. \ref{fig:AT2021lo_4FGL}. These monthly binned light curves were obtained from the Fermi LAT Light Curve Repository\footnote{\url{https://fermi.gsfc.nasa.gov/ssc/data/access/lat/LightCurveRepository/}} \citep[LCR;][]{Abdollahi2023ApJSFermiLAT}.

\subsection{Spatial associations}

In addition to the spatio-temporal associations, we investigated the spatial associations of TDEs and neutrinos. We find 26 spatial associations (not counting TDEs AT2019dsg/IC191001A and AT2021lo/IC220304A). The results of the spatial associations are presented in Table \ref{tab:spatial}. Most neutrinos arrive either much earlier or much later than the TDE flare, except for TDE AT2020acka (see Sect. \ref{subsubsect:TDE2020acka}). Furthermore, we repeated the spatial analysis for the 4FGL sources for the neutrinos in Table \ref{tab:spatial}, with the number of spatially associated 4FGL sources presented in the final column. 

We note that TDEs AT2017eqx, AT2020mot, and AT2019azh are each spatially associated with more than one neutrino (two neutrinos for the former two, three neutrinos for the latter), while neutrinos IC140223A, IC140410A, IC150224A, IC221210A, and IC230707B are spatially associated with four, two, three, two and three TDEs respectively. However, as seen in the fifth column of Tab. \ref{tab:spatial}, all aforementioned neutrino events have very large 90\% error areas, resulting in more events falling within them. Moreover, \citet{Liodakis2022A&AspatialFutile} demonstrated that the connection between neutrinos and AGN cannot be established on spatial association alone (see their Sect. 3). Similarly for TDEs, spatial association alone lacks the sufficient statistical power to draw meaningful conclusions.

\subsubsection{TDE AT2020acka/IC200410A}\label{subsubsect:TDE2020acka}

\begin{figure}[!htbp]
   \resizebox{\hsize}{!}
    {\includegraphics[]{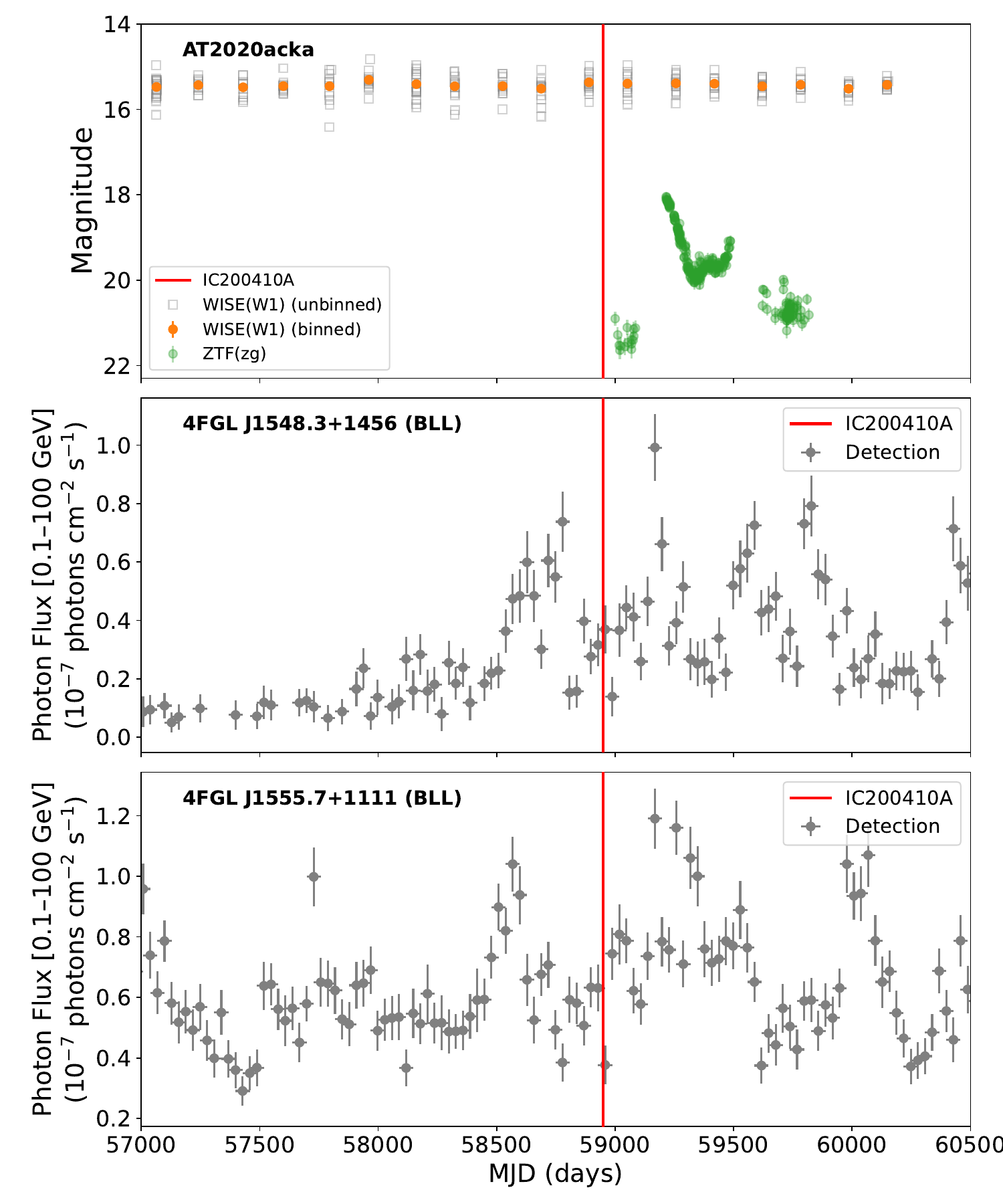}}
    \caption{Light curves for TDE AT2020acka (optical and infrared; top), 4FGL J1548.3+1456 ($\gamma$-ray observations in monthly bins; middle) and 4FGL J1555.7+1111 ($\gamma$-ray observations in monthly bins; bottom). The red vertical line indicates the neutrino arrival time of IC200410A.}
    \label{fig:bl lac}
\end{figure}

Although spatial association alone is not enough to draw a conclusion, we investigated this source and neutrino pair, which shows a spatial association only 50 days before the chosen TDE flare range (upper panel of Fig. \ref{fig:bl lac}). However, the neutrino 90\% confidence area in the sky contains 35 4FGL sources: that 15 BLL, 11 BCU, one RDG, two FSRQ, and six unclassified.

We find that 4FGL J1548.3+1456 (BLL) and 4FGL J1555.7+1111 (BLL) were flaring in $\gamma$ rays at the neutrino arrival time (middle and bottom panels of Fig. \ref{fig:bl lac}). The monthly binned light curves were again obtained from the Fermi LAT LCR. This behaviour resembles that of blazar TXS 0506+056, where the neutrino arrival coincided with a $\gamma$-ray flaring state \citep{IceCube2018SciBlazar}; hence, neutrino event IC200410A could be produced by 4FGL J1548.3+1456 or 4FGL J1555.7+1111. Interestingly, the latter appears also in a TeV source catalogue \citep[TeVCat\footnote{\url{https://tevcat2.tevcat.org/}};][]{Wakely2008ICRCTeVCat}, as TeV J1555+111. Moreover, we note that the signalness of this neutrino event is 0.305 (i.e. only a 30.5\% chance of being astrophysical), making it more likely to be atmospheric.

\subsection{Jetted TDEs}

\begin{figure}[!htbp]
   \resizebox{\hsize}{!}
    {\includegraphics[]{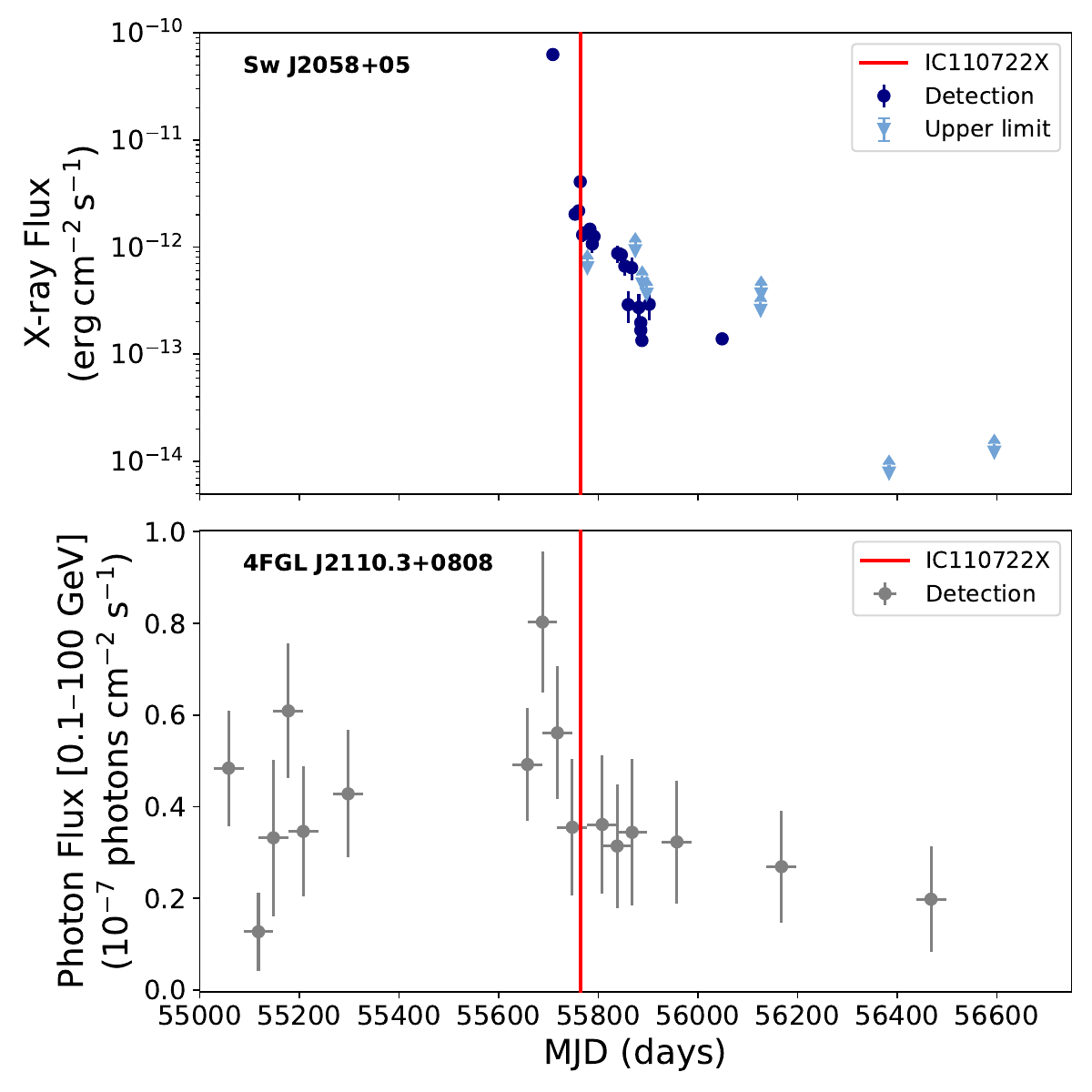}}
    \caption{Light curves of Sw J2058+05 (unbinned X-ray data; upper panel) and 4FGL J2110.3+0808 ($\gamma$-ray observations in monthly bins; bottom panel). The  vertical red line indicates the neutrino arrival time of IC110722X.}
    \label{fig:fsqr}
\end{figure}

We also investigated the association of the jetted TDEs in the sample of \citet{Langis2025arXivTDECat} within the neutrino time range, namely Swift J164449.3+573451, Swift J2058.4+0516 (hereafter Sw J1644+47 and Sw J2058+05, respectively), and AT2022cmc. We applied the spatio-temporal method (without performing the MC experiment to retrieve the randomized AI TS distribution) and find that Sw J2058+05 is spatio-temporally associated with neutrino event IC110722X (named by \citetalias{Kouch2024BlazarNeutrinoAssociation}), yielding $AI_{raw}=1$ and $AI_{weighted} \approx 0.058$. The neutrino arrived $\sim 60$ days after the first X-ray observation (upper panel of Fig. \ref{fig:fsqr}) and has a significant chance of being astrophysical with a signalness of 0.56. 

The 90\% confidence area of neutrino event IC110722X ($\Omega_{90}=63.5$ deg$^2$) includes ten 4FGL sources, four of which are classified as BLL (4FGL J2049.9+1002, 4GL J2050.0+0408, 4FGL J2052.5+0810, and 4FGL J2109.6+0440), four as BCU (4FGL J2101.3+0912, 4FGL J2103.0+0342, 4FGL J2104.7+0108, and 4FGL J2110.3+0404), one as FSRQ (4FGL J2110.3+0808) and one source is unclassified (4FGL J2058.6+0550).

Interestingly, 4FGL J2110.3+0808 (an FSRQ) appears to be in the end of a $\gamma$-ray flaring episode at the neutrino arrival time (bottom panel of Fig. \ref{fig:fsqr}). Even though the spotlight has been shone on BLL objects for neutrino production, FSRQs have also been studied recently \citep{Rodrigues2018ApJFSRQS, Righi2020A&AFSRQS}. However, \citet{Righi2020A&AFSRQS} find that the expected HE neutrinos from FSRQs are in the sub-EeV to EeV range, making them undetectable with current technology. Additionally, \citet{Liodakis2020ApJCrisis} studied the energetic requirements for 145 $\gamma$-ray blazars for the proton synchrotron model \citep[e.g.][]{Aharonian2000NewAProtonSynchrotron} and find that in all cases, the neutrino emission peaked at $\sim0.1-1$ EeV as well.

We can apply the same analysis as at the beginning of this section \citep[inspired by][]{Bartos2021ApJPieChart} to the jetted TDEs. We find that $f_{jetted\ TDE}\approx 0.013=1.3\%$, where all the input parameters are the same as Eq. \ref{eq:fraction} (with $f_C=0.5$), except $N_{det}=1$. This result is consistent with \citet{Stein2019ICRCrefereepaper}, who found that jetted TDEs contribute less than $1.3\%$ of the neutrino flux. We note that this is a rough estimate, where the completeness factor of jetted TDEs could be significantly overestimated.

\section{Conclusions}\label{sec:Conclusions}

We used the neutrino sample of \citetalias{Kouch2024BlazarNeutrinoAssociation} to investigate a potential correlation between IC neutrino events and optically selected TDEs. We also generated two different types of mock data to simulate two distinct cases: fully correlated TDE-neutrino data and completely random data. We applied the spatio-temporal method both to observed and simulated data to test whether an association between TDEs and neutrinos exists. Our key findings are as follows:

\begin{enumerate}
    \item We retrieved the association of TDE AT2019dsg with neutrino event IC191001A, as expected from the literature \citep{Stein2021NatAsAT2019dsg}, as well as a new association of TDE AT2021lo with IC220304A. The neutrino arrives after the optical peak in both cases; however, as shown from our simulations, this is the most probable scenario by random chance. Therefore no conclusion can be drawn regarding the physical origin of a potential delayed neutrino emission. Additionally, there is a blazar candidate inside the error region of IC191001A, observed to be flaring in the optical, meaning that the neutrino could be produced from that source. Furthermore, IC220304A is spatially coincident with three $\gamma$-ray sources flaring around the neutrino arrival time.
    \item We do not find a statistically significant association between our TDE sample and the HE neutrino sample. Through simulations we discovered that in order to obtain a $4\sigma$ signal we would require at least six or nine observed TDE-neutrino associations for the weighted or raw cases, respectively.
    \item We applied the spatio-temporal method to three jetted TDEs and find Sw J2058+05 to be associated with neutrino event IC110722X. The neutrino event arrives after the peak of the X-ray light curve. However, an FSRQ source also lies within the error region of IC110722X, which is in the end of a $\gamma$-ray flare during the neutrino arrival time.
\end{enumerate}

Although our results show no association between the optical TDEs and neutrino events in our samples, this correlation should be further studied in future work (X-ray-selected or infrared-selected samples). Upcoming surveys such as the Legacy Survey of Space and Time\footnote{\url{https://www.lsst.org/about}} \citep[LSST; ][]{Ivezic2008LSST} will push the boundaries of transient astronomy. These surveys along with next-generation neutrino observatories such as KM3NeT, IceCube-Gen2\footnote{\url{https://www.icecube-gen2.de/index_eng.html}} \citep{Aartsen2021JPhGICGen2}, and Baikal-GVD \citep{Shoibonov2019JPhCSBAIKAL} will enable tests of this correlation with much larger and more representative optical TDE and neutrino event samples.

\begin{acknowledgements}
The authors thank the anonymous referee for comments and suggestions that helped improve this work. DAL and IL were funded by the European Union ERC-2022-STG - BOOTES - 101076343. Views and opinions expressed are however those of the authors only and do not necessarily reflect those of the European Union or the European Research Council Executive Agency. Neither the European Union nor the granting authority can be held responsible for them. KIIK has received funding from the European Research Council (ERC) under the European Union’s Horizon 2020 research and innovation programme (grant agreement No. 101002352, PI: M. Linares). PMK was supported by Academy of Finland projects 346071 and 345899. Computations for this paper were partially conducted on the Metropolis cluster, supported by the Institute of Theoretical and Computational Physics at the Department of Physics, University of Crete.
\end{acknowledgements}

\bibliographystyle{aa}
\bibliography{references.bib}

\begin{appendix}

\onecolumn\section{Badly sampled TDE light curves}\label{Appendix:durations}

In our TDE sample, eight sources (AT2018dyb, AT2018bsi, AT2017eqx, iPTF16axa, iPTF15af, ASASSN-15oi, LSQ12dyw, and PS1-10jh) have badly sampled light curves, making the estimation of the duration unreliable using the Bayesian Block algorithm. Below we summarize the process to determine their duration:

\begin{itemize}
    \item AT2018dyb: The decay time can be calculated using the Bayesian Block algorithmn, however the rise time does not have data prior to the start of the rise. In order to be sure that the whole flare is considered, we calculate the mean sampling rate of the flare and then subtract three times the mean sampling rate from the estimated Bayesian Block start time (58297.0325 MJD). The start of the duration for AT2018dyb is taken to be 58271.547 MJD.
    \item AT2018bsi: Similar to AT2018dyb, the decay time can be calculated. However, there are only data near the peak for the rise time. For this reason, we use the ODR best fit relation from \citet{Langis2025arXivTDECat}, plugging in the decay time and the decay error, and retrieving the rise time and rise error. We then consider the duration as described in Sect. \ref{ssec:TDEs}.
    \item AT2017eqx: For this TDE, the light curves were the worst sampled. There are only 2 points before the peak (from different bands), the peak is not easily determined and the decay stops before the plateau. The data are also spread between 5 bands, $g,\ r,\ i,\ z$ and $o$. For this reason, we combine the three light curves with the most points ($g,\ i,\ o$) and determine the peak as the average of the two lowest magnitude points, using as the error their standard deviation. We then calculate the mean sampling rate of this combined light curve, multiply it by 3 and add it to the last observation, giving us the end date of the flare (58028.100 MJD). Similar to AT2018bsi, we estimate the rise time using the best fit ODR relations from \citet{Langis2025arXivTDECat} and consider the duration as described in Sect. \ref{ssec:TDEs}.
    \item iPTF16axa: For this TDE, only the decay part of the flare is observed. First we estimate the mean sampling rate of the light curve. Then, we consider as the peak, the first observation ($g$-band light curve from TDECat) and as the end of the flare the final observation, with error the mean sampling rate for both. Similar to AT2017eqx we calculate the rise time, rise error and the duration.
    \item iPTF15af: The rise time for this TDE is well sampled and the decay is not. We first estimate the mean sampling rate. We use as the start time the first point of the $r-$band light curve on TDECat and the mean sampling rate as the error. Since the peak is well sampled, we consider as the peak the minimum magnitude point and as the error the mean sampling rate. We now use the difference of the peak and the start time as the rise time and the error of the two taken in quadrature as the rise error. We use the ODR best fit relation from Langis et al. 2025 to calculate the decay time and decay error. The duration is determined as described in Sect. \ref{ssec:TDEs}.
    \item ASASSN-15oi: Similar methodology to iPTF16axa.
    \item LSQ12dyw: The light curve has a lot of noise, hence we manually selected a duration that included the whole flaring part of the light curve, from 55700 MJD to 56800 MJD.
    \item PS1-10jh: We use as the duration the whole PS1-$g$ light curve that is available on TDECat. 
\end{itemize}

\onecolumn\section{Spatially associated TDEs and neutrinos}\label{Appendix:Table_1}

\begin{table*}[htbp!] 
    \centering 
    \caption{Spatial TDE-neutrino associations} 
    \label{tab:spatial}
    \begin{tabular}{l l c c c c}
        \toprule 
        TDE name    & Neutrino name& \begin{tabular}[c]{@{}c@{}}Arrival\\ (days) \end{tabular} & Signalness & \begin{tabular}[c]{@{}c@{}}$\Omega_{90}$\\ (deg$^2$) \end{tabular}  & \begin{tabular}[c]{@{}c@{}}4FGL sources\\ in $\Omega_{90}$ \end{tabular} \\[3pt]
        \midrule 
        PTF09ge & IC140324A & 1833 & 0.403 & 55.68 & 5 \\
        PS1-10jh & IC150224A & 1751 & 0.379 & 52.96 & 6 \\
        iPTF15af & IC230707B & 3121 & 0.466 & 295.41 & 31 \\
        AT2017eqx & IC221210A & 2012 & 0.508 & 217.34 & 27 \\
        AT2017eqx & IC210608A & 1462 & 0.315 & 89.63 & 12 \\
        AT2018zr & IC140223A & -1400 & 0.430 & 289.19 & 39 \\
        AT2018jbv & IC110907A & -2436 & 0.506 & 13.51 & 1 \\
        AT2019azh & IC230217A & 1529 & 0.454 & 19.51 & 2 \\
        AT2019azh & IC230707B & 1668 & 0.466 & 295.41 & 31 \\
        AT2019azh & IC161021A & -782 & 0.434 & 17.19 & 1 \\
        AT2019mha & IC150224A & -1529 & 0.379 & 52.96 & 6 \\
        AT2020mot & IC140410A & -2093 & 0.627 & 2469.15 & 54 \\
        AT2020mot & IC190629A & -186 & 0.343 & 1126.63 & 17 \\
        AT2020yue & IC130408A & -2510 & 0.525 & 11.77 & 2 \\
        AT2021ack & IC220928A & 885 & 0.378 & 5.53 & 2 \\
        AT2020acka & IC200410A & -50 & 0.305 & 224.63 & 35 \\
        AT2021nwa & IC150224A & -2222 & 0.379 & 52.96 & 6 \\
        AT2021yte & IC180807A & -964 & 0.276 & 36.25 & 4 \\
        AT2021utq & IC161127A & -1682 & 0.453 & 808.46 & 47 \\
        AT2022aee & IC140410A & -2713 & 0.627 & 2469.15 & 54 \\
        AT2022csn & IC140223A & -2789 & 0.430 & 289.19 & 39 \\
        AT2022gri & IC140223A & -2789 & 0.430 & 289.19 & 39 \\
        AT2022bdw & IC230707B & 600 & 0.466 & 295.41 & 31 \\
        AT2023kvy & IC150129A & -2853 & 0.334 & 56.11 & 9 \\
        AT2023ugy & IC221210A & -248 & 0.508 & 217.34 & 27 \\
        AT2023xen & IC140223A & -3482 & 0.430 & 289.19 & 39 \\
        \bottomrule 
        \hspace{0.1cm}
    \end{tabular}
    \caption*{{ Notes:} The first two columns list the TDE and the spatially associated neutrino names respectively. The third column indicates the neutrino arrival time, expressed in days before or after the TDE flare duration considered ("$-$" before, "$+$" after). The fourth and fifth columns present the neutrino signalness and 90\% sky error area in deg$^2$. The final column shows the number of 4FGL sources within the neutrino's 90\% confidence region on the sky.} 
\end{table*}

\end{appendix}
\end{document}